\documentclass[11pt]{iopart}

\usepackage[british]{babel}
\usepackage{color}
\usepackage{mathrsfs}
\usepackage{setstack}
\usepackage{iopams}
\usepackage{bm}
\usepackage{graphicx}

\newcommand{\erf}{\mathrm{erf}}

\begin{document}

\title[]{First passage statistics for diffusing diffusivity}

\author{Vittoria Sposini$^{\dagger,\ddagger}$, Aleksei Chechkin$^{\dagger,
\flat}$ \& Ralf Metzler$^{\dagger}$}
\address{$\dagger$ Institute for Physics \& Astronomy, University of Potsdam,
14476 Potsdam-Golm, Germany}
\address{$\ddagger$ Basque Center for Applied Mathematics, 48009 Bilbao, Spain}
\address{$\flat$ Akhiezer Institute for Theoretical Physics, 61108 Kharkov, Ukraine}
\ead{rmetzler@uni-potsdam.de (Corresponding author: Ralf Metzler)}

\begin{abstract}
A rapidly increasing number of systems is identified in which the stochastic
motion of tracer particles follows the Brownian law $\langle\mathbf{r}^2(t)
\rangle\simeq Dt$ yet the distribution of particle displacements is strongly
non-Gaussian. A central approach to describe this effect is the diffusing
diffusivity (DD) model in which the diffusion coefficient itself is a stochastic
quantity, mimicking heterogeneities of the environment encountered by the
tracer particle on its path. We here quantify in terms of analytical and
numerical approaches the first passage behaviour of the DD model. We observe
significant modifications compared to Brownian-Gaussian diffusion, in particular
that the DD model may have a more efficient first passage dynamics. Moreover we
find a universal crossover point of the survival probability independent of the
initial condition.
\end{abstract}

\section{Introduction}

Since its original systematic study 190 years ago by Robert Brown \cite{Brown},
diffusion of molecular and (sub-)micron-sized entities has been identified as
the dominant form of thermally driven, passive transport in numerous biological
and inanimate systems. The two hallmark features of diffusion is the linear
growth $\langle\mathbf{r}^2(t)\rangle=2Ddt$ of the mean squared displacement
(MSD) with diffusion coefficient $D$ in $d$ spatial dimensions, and the Gaussian
distribution of displacements \cite{vankampen}. With increasing complexity of
the studied systems deviations from these two central properties have been
unveiled over the years. Thus, anomalous diffusion with an MSD of the form
$\langle\mathbf{r}^2(t)\rangle\simeq t^{\alpha}$ was observed in a large
range of systems \cite{hoefling,lene_rev}. Along with such observations a rich
variety of generalised stochastic processes has been developed \cite{bouchaud,
pccp}. The displacement distribution of anomalous diffusion processes may be
inherently Gaussian (such as for fractional Brownian motion \cite{fBM}) or
non-Gaussian (for instance, for processes characterised by scale-free trapping
time distributions \cite{scher} or space-dependent diffusivity models
\cite{andrey}).

Recently a lage variety of systems have been reported in which the MSD exhibits
the linear growth in time $\langle\mathbf{r}^2(t)\rangle\simeq Dt$ of Brownian
(Fickian) transport, however, the distribution of displacements $P(\mathbf{r},t)$
is pronouncedly non-Gaussian \cite{Wang:BYNG2}. Pertinent examples include the
motion of tracer beads along tubular or membrane structures or in gels and
colloidal suspensions \cite{Wang:BYNG2,Wang:BYNG3,Goldstein:BYNG9}, and
the motion of nematodes \cite{hapca}
and single cells on substrates \cite{beta}. As long as the displacement
distribution $P(\mathbf{r},t)$ has a fixed shape for any times $t$, one possible way
to model the non-Gaussianity is the concept of superstatistics \cite{Superst_1,
Superst_2} which introduces a distribution $p_D(D)$ of the diffusion coefficient
and then averages individual Gaussian distributions with one given $D$ value
over this $p_D(D)$. However, this approach does not work when eventually a
crossover to an effective Gaussian is observed \cite{Wang:BYNG2,Wang:BYNG3}. For
the latter case Chubynsky and Slater introduced the diffusing diffusivity (DD) model
\cite{Chubynsky:DD1}, see also \cite{Chubynsky_new:DD1,Sebastian:DD3,
Sebastian:DD3_bis,tyagi:DD5,Chechkin:DD2,Sposini:DD2,Grebenkov:DD4}: In this
popular approach the diffusion coefficient is assumed to be a stochastic variable
itself, described by a stationary process. Consequently the system is initially
described by a non-Gaussian displacement distribution. Beyond a characteristic
time scale a crossover occurs to a Gaussian behaviour characterised by an effective
value of the diffusivity.

We here study the first passage behaviour of the DD model.
The concept of first passage is ubiquitously used in statistical physics and
its applications, for instance, to quantify when a diffusing particle reaches
a reaction centre or a stochastic process exceeds a given threshold value
\cite{redner,ralf_gleb_sid_book}. Based on the minimal model for DD
\cite{Chechkin:DD2} we derive the first passage
behaviour in both semi-infinite and finite systems. We find that the DD
dynamics may outperform Brownian-Gaussian normal diffusion at
intermittent times in a semi-infinite domain while the long time behaviour
matches exactly the Brownian-Gaussian result with an effective diffusivity.
We also observe an interesting universal crossover point of the survival probability
which is independent of the initial particle position.
In finite domains the mean first passage time of the DD model is longer than
in the Brownian-Gaussian case. Concurrently, in the DD model
the divergence of the mean first passage
time observed in the superstatistical approach is rectified.

In section 2 we briefly recall the basic properties of the minimal diffusing
diffusivity model \cite{Chechkin:DD2}. The survival probabilities for the
semi-infinite and finite domains are then derived in section 3 along with
their short and long time asymptotes. Section 4 provides a detailed
discussion of the results including a relation to the standard Brownian-Gaussian
first passage behaviour. A short conclusion is presented in section 5.

\section{Minimal model for Brownian yet non-Gaussian diffusion}

The model we study is the so-called minimal diffusing diffusivity (DD) model which was 
introduced to describe diffusion in heterogeneous environments \cite{Chechkin:DD2}. 
In this model the diffusivity is defined as a stochastic process itself, in terms
of the squared Ornstein-Uhlenbeck process, guaranteeing the stationarity of $D(t)$.
In dimensionless units the minimal DD model is defined by the set of Langevin equations
\cite{Chechkin:DD2}
\begin{eqnarray}
\frac{d}{dt}\mathbf{r}(t)&=&\sqrt{2D(t)}{\bm \xi}(t)\nonumber\\
D(t)&=& \mathbf{Y}^2(t),\quad\frac{d}{dt}\mathbf{Y}(t)=-\mathbf{Y}+{\bm\eta}(t),
\label{eqn:XDD}
\end{eqnarray}
where the components of $\bm\xi(t)$ and $\bm\eta(t)$ are independent white Gaussian
noises and $\mathbf{Y}$ represents an $d$-dimensional Ornstein-Uhlenbeck process. The
dimensionless Ornstein-Uhlenbeck process here has a characteristic crossover time of
unity. We assume the diffusivity to start from equilibrium initial conditions (the
non-equilibrium case is discussed in \cite{Sposini:DD2}). This leads to the
superstatistical short time diffusivity distribution
\begin{equation}
p_D(D)=\left\{\begin{array}{lr}(\sqrt{\pi D})^{-1}\e^{-D}, & d=1\\
\e^{-D}, & d=2\\
(2\sqrt{D/\pi})\e^{-D}, & d=3 
\end{array}\right.,
\label{eqn:PD}
\end{equation}
for 1, 2 and 3 dimensions. While the dominating exponential tail is common to all
$d$, there is a pole at $D\to0$ in $d=1$ \cite{Chechkin:DD2}. As we
showed previously, the minimal DD model can be written using the concepts of
subordination \cite{Subord} through the relation \cite{Chechkin:DD2}
\begin{equation}
P(\mathbf{r},t|\mathbf{r}_0)=\int_0^{\infty}G(\mathbf{r},\tau|\mathbf{r}_0,D=1)
T_d(\tau,t)d\tau,
\label{eqn:PDD}
\end{equation}
of the probability density function (PDF) $P(\mathbf{r},t|\mathbf{r}_0)$ of
displacement and the Gaussian
\begin{equation}
G(\mathbf{r},t|\mathbf{r}_0,D)=(4 \pi D t)^{-d/2}\exp\left((\mathbf{r}-\mathbf{r}_0)^2
/[4Dt]\right)
\end{equation}
with fixed diffusion coefficient $D$. The subordinator $T_d(\tau,t)$ represents the PDF
of the process $\tau(t)=\int_0^t\mathbf{Y}^2(t')dt'$ and is defined through its Laplace
transform \cite{Chechkin:DD2}
\begin{equation}
\fl\tilde{T}_d(s,t)=\exp(dt/2)\left[\frac{1}{2}(\sqrt{1+2s}+\frac{1}{\sqrt{1+2s}})
\sinh \left(t \sqrt{1+2s}\right)+\cosh \left(t \sqrt{1+2s}\right)\right]^{-d/2}
\label{eqn:T}
\end{equation} 
with short and long times limits
\begin{eqnarray}
\fl\tilde{T}_d(s,t)&\sim t^{-d/2}\left(s+1/t\right)^{-d/2}, & t\ll 1,\label{eqn:TST}\\
\fl\tilde{T}_d(s,t)&\sim2^{d/2}\exp\left(\frac{dt}{2}(1-\sqrt{1+2s})\right)
\left(1+\frac{1}{2}\left(\sqrt{1+2s}+\frac{1}{\sqrt{1+2s}}\right)\right)^{-d/2}, &
t\gg 1.
\label{eqn:TLT}
\end{eqnarray}
At short times the diffusivity varies slowly and we can assume it to be almost
constant. In this limit the DD model thus reduces to the superstatistical
approximation of the DD model in which each particle has a constant random diffusion
coefficient with distribution $p_D(D)$ \cite{Superst_1,Superst_2}: on the ensemble
level this implies that the PDF can be written as $P_\mathrm{sup}(\mathbf{r},t|
\mathbf{r}_0)=\int_0^\infty G(\mathbf{r},t|\mathbf{r}_0,D)p_D(D)dD$, such that the
short time PDF explicitly reads
\begin{equation}
P_\mathrm{ST}(\mathbf{r},t|\mathbf{r}_0)=P_\mathrm{sup}(\mathbf{r},t|\mathbf{r}_0) =
\left\{\begin{array}{lr}(\pi t^{1/2})^{-1}K_0\left(|x-x_0|/t^{1/2}\right), & d=1\\
(2\pi t)^{-1}K_0\left(|r-r_0|/t^{1/2}\right), & d=2\\
(2\pi^2 t^{3/2})^{-1}K_0\left(|r-r_0|/t^{1/2}\right), & d=3
\end{array}
\right.
\label{eqn:PDDST}
\end{equation} 
where $K_0(x)$ is a modified Bessel function of the second kind with exponential
asymptote $K_0(z)\sim\sqrt{\pi/(2z)}e^{-z}$ \cite{prudnikov2}. At long times
the DD process crosses over to a purely Gaussian process with
PDF $ P_\mathrm{LT}(\mathbf{r},t|\mathbf{r}_0)=G(\mathbf{r},t|\mathbf{r}_0,D=
\langle D\rangle_\mathrm{st})$ with the stationary diffusivity $\langle
D\rangle_\mathrm{st}=d/2$ \cite{Chechkin:DD2}. For all $t$ the MSD is given by
$\langle(\mathbf{r}(t)-\mathbf{r}_0)^2\rangle=2d\langle D \rangle_\mathrm{st}t$.

We showed in \cite{Sposini:DD2} that there is a stochastic counterpart to this
superstatistical approximation, defined through the generalised grey Brownian
motion (ggBM) formalism \cite{Mura:ggBM}, $\mathbf{r}(t)=\sqrt{2D}\times\mathbf{
W}(t),$ where $D$ is the random and constant diffusion coefficient and $\mathbf{
W}(t)$ is the $d$-dimensional Wiener process or standard Brownian motion. Note
that while the DD model represents the heterogeneity of the medium in some mean
field sense \cite{Chechkin:DD2} the ggBM model describes an heterogeneous
ensemble of particles \cite{spos-pagni:2} .

\section{Results for the survival probabilities}

The first passage time PDF of a stochastic process is the negative time
derivative of the survival probability, $\wp(t)=-d S(t)/dt$. We here obtain
the survival probability for semi-infinite and finite domains using the
above subordination relation.

\subsection{Survival of diffusing diffusivity model in semi-infinite domain}

We begin our study with the semi-infinite interval $d=1$. Following the approach
for standard diffusion \cite{redner} we use the method of images for the initial
particle position $x_0$. Combined with the subordination principle (\ref{eqn:PDD})
we get the image propagator
\begin{equation}
\fl P(x,t|x_0)=\int_0^{\infty}\left(G(x,\tau|x_0,D=1)-G(x,\tau|-x_0,D=1)\right)T_1(
\tau,t)d\tau.
\end{equation}
After Fourier transform we obtain
\begin{equation}
\fl\hat{P}(k,t|x_0)=\int_0^\infty T_1(\tau,t)e^{-k^2\tau}\left(e^{ikx_0}-e^{-ikx_0}
\right)d\tau=(e^{ikx_0}-e^{-ikx_0})\tilde{T}_1(s=k^2,t).
\end{equation}
Here $\hat{\cdot}$ and $\tilde{\cdot}$ indicate the Fourier and Laplace transforms
of the functions, respectively. We then calculate the survival probability in the
semi-infinite domain,
\begin{equation}
S(t|x_0)=\int_0^\infty P(x,t|x_0)dx=\int_0^\infty dx\int_{-\infty}^{+\infty} 
\frac{dk}{2\pi}e^{-ikx}\hat{P}(k,t|x_0).
\label{eqn:SDDint}
\end{equation}
To check normalisation, we first see from expression (\ref{eqn:T}) that $\tilde{
T}_1(s,0)=1$. Then,
\begin{eqnarray}
S(0|x_0)=\int_0^\infty dx\int_{-\infty}^{+\infty}\frac{dk}{2\pi}(e^{-ik(x-x_0)}
-e^{-ik(x+x_0)})=1,
\end{eqnarray}
where we used that $\int_{-\infty}^{\infty}dk/(2\pi)\exp(-ikx)=\delta(x)$. Moreover,
plugging the long time limit for $\tilde{T}_1(s=k^2,t)$ in (\ref{eqn:TLT}) into the
expression for $S(t|x_0)$ one can readily show that $S(t\to\infty|x_0)=0$, as it
should.

The direct calculation of the integral (\ref{eqn:SDDint}) is not easy to perform,
we here focus on the short and long time regimes. At short times, $\tilde{T}_1(s
=k^2,t)$ is given by (\ref{eqn:TST}) and thus
\begin{eqnarray}
\fl S_\mathrm{ST}(t|x_0)&=&\int_0^{\infty}dx\int_{-\infty}^{\infty}\frac{dk}{2\pi}
e^{-ikx}(e^{ikx_0}-e^{-ikx_0})\frac{t^{-1/2}}{\sqrt{k^2 +1/t}}\nonumber\\
\fl &=&\frac{1}{2\pi\sqrt{t}}\left(\int_0^\infty dx\int_{-\infty}^{\infty}dk 
\frac{e^{-ik(x-x_0)}}{\sqrt{k^2 +1/t}}-\int_0^\infty dx \int_{-\infty}^{\infty}dk 
\frac{e^{-ik(x+x_0)}}{\sqrt{k^2 +1/t}}\right)\nonumber\\
\fl &=& \frac{1}{\pi \sqrt{t}} \int_0^\infty \left[ K_0\left(\frac{|x-x_0|}{\sqrt{
t}}\right)-K_0\left(\frac{|x+x_0|}{\sqrt{t}}\right)\right]dx,
\label{DD=ggBM}
\end{eqnarray}
Splitting the integral and changing variables we obtain
\begin{eqnarray}
\fl S_\mathrm{ST}(t|x_0)&=&\frac{1}{\pi\sqrt{t}}\left[\int_0^{x_0}K_0\left(\frac{
x_0-x}{\sqrt{t}}\right)dx+\int_{x_0}^\infty K_0\left(\frac{x-x_0}{\sqrt{t}}\right)dx 
-\int_0^\infty K_0\left(\frac{x+x_0}{\sqrt{t}}\right)dx\right]\nonumber\\
\fl &=&\frac{2}{\pi}\int_0^{x_0/\sqrt{t}}K_0(z)dz.
\end{eqnarray}
Using $\int_0^a K_0(z)dz=a\pi/2\left(K_0(a)L_{-1}(a)+K_1(a)L_0(a)\right)$, with the
modified Struve function $L_\nu(z)$ \cite{prudnikov2},
\begin{equation}
S_\mathrm{ST}(t|x_0)=\frac{x_0}{\sqrt{t}}\left[K_0\left(\frac{x_0}{\sqrt{t}}\right)
L_{-1}\left(\frac{x_0}{\sqrt{t}}\right)+K_1\left(\frac{x_0}{\sqrt{t}}\right)L_0\left(
\frac{x_0}{\sqrt{t}}\right)\right].
\label{eqn:S_DD_ST}
\end{equation}
The same result can be obtained both inserting directly the short times approximation
(\ref{eqn:PDDST}) of the propagator in the result of the images method of images and
calculating directly the superstatistical integral valid for the survival probability.

At long times, when in equation (\ref{eqn:TLT}) we only consider the tails of the
distribution, $\tilde{T}_1(s=k^2,t)\sim\exp\left(-k^2t/2\right)$. This approximation
leads to 
\begin{eqnarray}
\fl S_\mathrm{LT} (t|x_0)&=&\int_0^\infty dx\left[\int_{-\infty}^{\infty}\frac{dk}{2
\pi}\exp\left(-ik (x-x_0)-\frac{k^2t}{2} \right)-\exp\left(-ik (x+x_0)-\frac{k^2t}{2}
\right)\right]\nonumber\\
\fl &=& \frac{1}{\sqrt{2 \pi t}} \int_0^\infty dx \left[ \exp\left(-\frac{(x-x_0)^2}{2t}\right) - 
\exp\left(-\frac{(x+x_0)^2}{2t}\right) \right]  \nonumber \\
\fl &=& \erf\left( \frac{x_0}{\sqrt{2 t}} \right)= \erf\left( \frac{x_0}
{\sqrt{4 \langle D \rangle_\mathrm{st} t}} \right) .
\label{S_DD_lt}
\end{eqnarray}
This result equals the one for Brownian diffusion in a semi-infinite domain, in
agreement with the fact that at long times the DD model shows a crossover to
Gaussian diffusion with effective diffusivity $ \langle D \rangle_\mathrm{st}$. 
In analogy with Brownian diffusion, this particularly leads to the divergence of
the mean first passage time, $\langle t\rangle=\infty$.

\subsection{Survival of diffusing diffusivity model in finite domain}

We now turn to a finite domain $[0,L]$ with absorbing boundaries at $x=0$
and $x=L$. Drawing on the subordination approach again, we map the images result
for the finite domain to obtain the DD propagator,
\begin{equation}
P(x,t|x_0)=\frac{2}{L}\sum_{n=1}^{\infty}\sin\left(\frac{\pi n}{L}x_0\right)\sin 
\left(\frac{\pi n}{L}x\right)\tilde{T}_1(\lambda_n^2,t).
\label{eqn:DDsubL}
\end{equation}
By integration we obtain the survival probability
\begin{equation}
S(t|x_0)=\frac{4}{\pi}\sum_{n=0}^{\infty}\sin\left(\frac{\pi(2n+1)}{L}x_0\right) 
\frac{\tilde{T}_1(\lambda_{2n+1}^2,t)}{(2n+1)},
\end{equation}
from which we obtain the limiting behaviours for short times,
\begin{equation}
S_\mathrm{ST} (t|x_0) \sim \frac{4}{\pi} \sum_{n=0}^{\infty} \sin 
\left( \frac{\pi (2n +1)}{L} x_0 \right) \frac{1}{(2n+1)\sqrt{\lambda_{2n+1}^2 t +1}},
\end{equation}
and for long times,
\begin{eqnarray}
\fl S_\mathrm{LT}(t|x_0)&\sim&\frac{4\sqrt{2}}{\pi} \sum_{n=0}^{\infty} \sin
\left( \frac{\pi (2n +1)}{L} x_0 \right)\exp\left(-\frac{t}{2}
\left[\sqrt{1+2\lambda_{2n+1}^2}-1\right]\right)\nonumber\\
\fl &&\times(2n+1)\left( 1+ \frac{1}{2}
\left(\sqrt{1+2\lambda_{2n+1}^2}+\frac{1}{\sqrt{1+2\lambda_{2n+1}^2}}
\right)\right)^{1/2}.
\label{eqn:SDDL_LT}
\end{eqnarray}
Note that, as in the previous case, the asymptotic behaviour at short times can also
be found through direct calculation of the superstatistical integral.

\section{Discussion of results}

\begin{figure}
\includegraphics[width=0.5\textwidth]{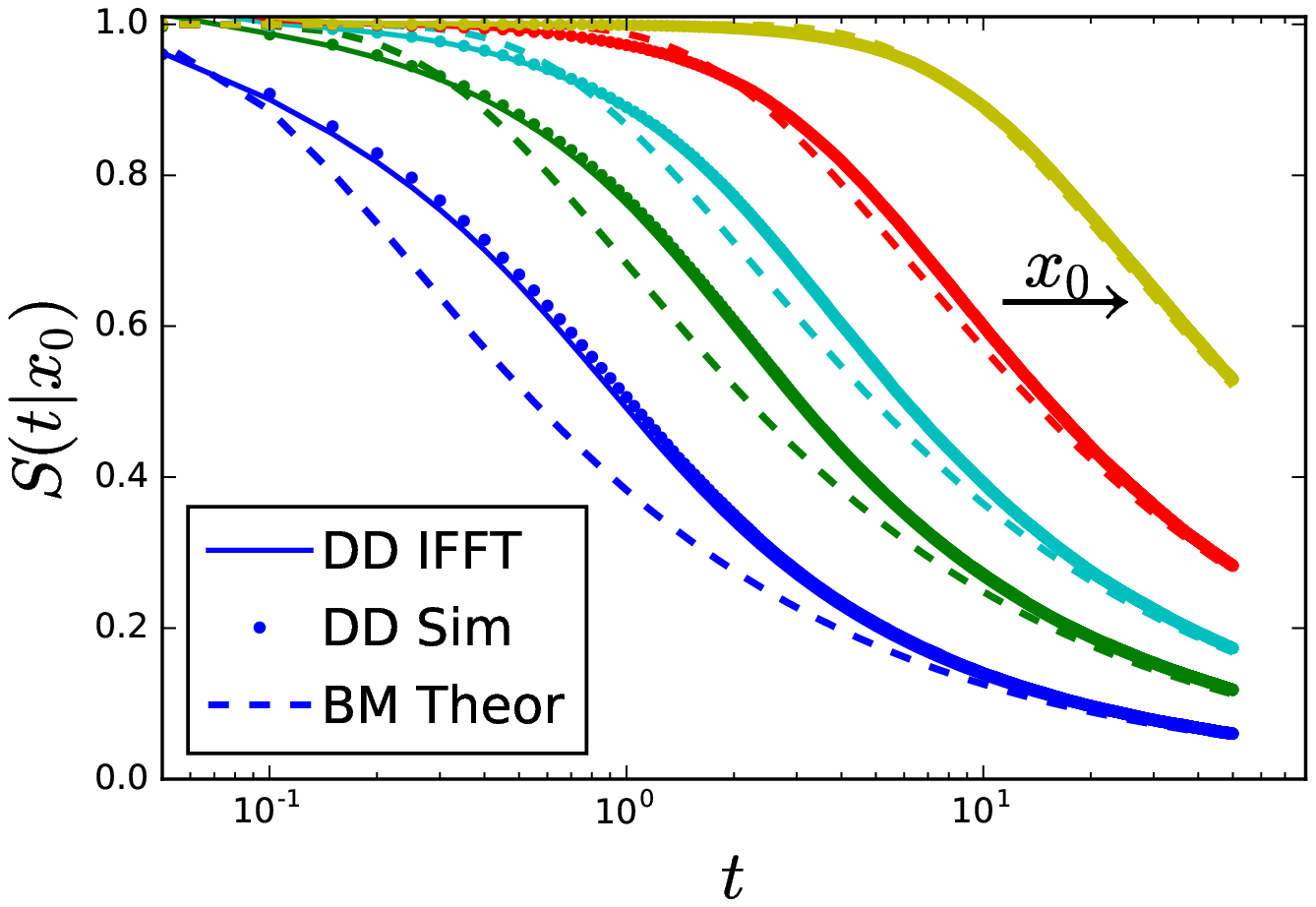}
\includegraphics[width=0.5\textwidth]{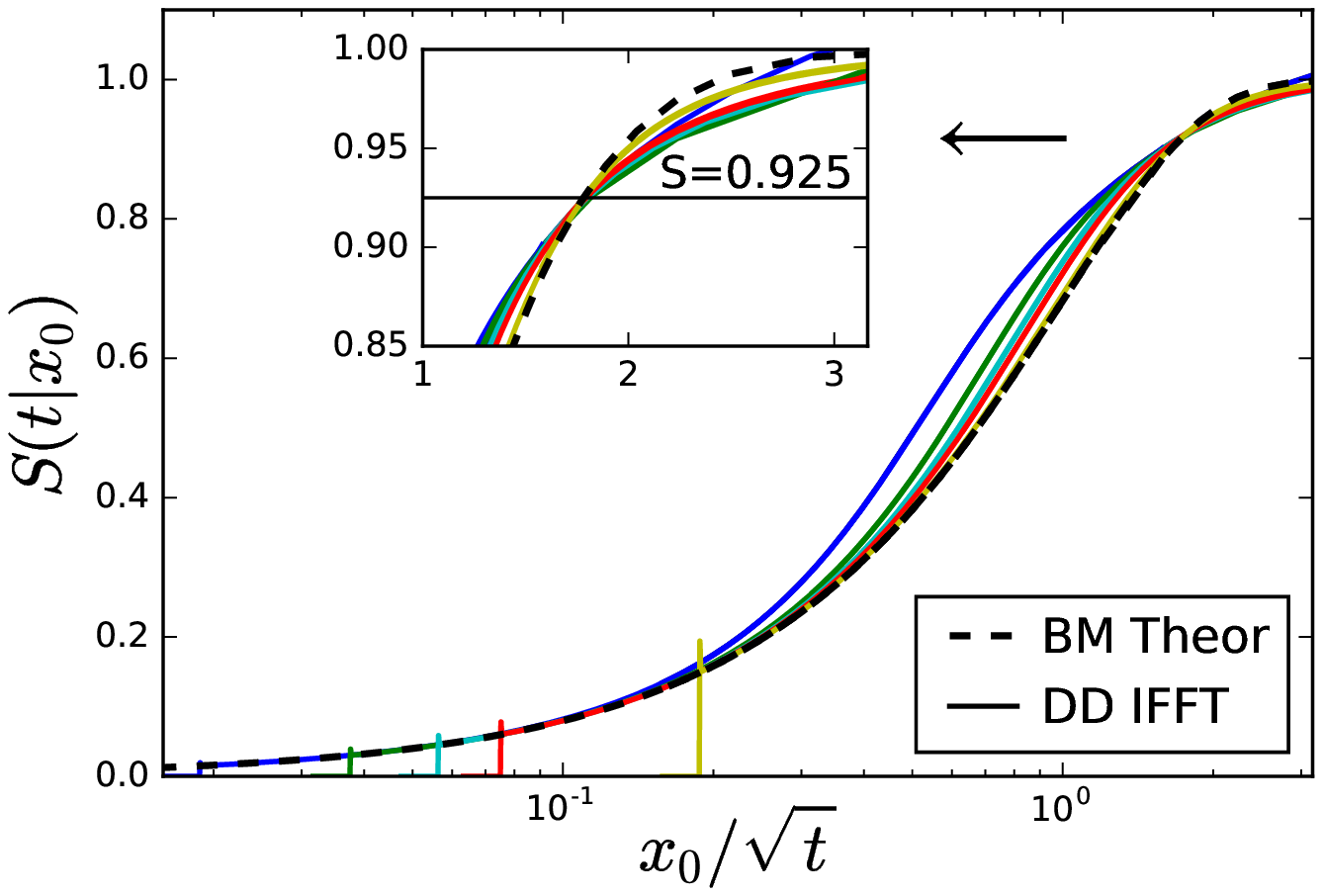}
\caption{Left: Comparison of numerical and analytical results for the survival
probability $S(t|x_0)$ in the semi-infinite interval. Different colours represent
the initial positions $x_0=0.5$, $1$, $1.5$, $2$, and $5$. Dashed lines in both
panels represent the results of the corresponding Brownian-Gaussian motion. The
numerical results (dots) obtained through Monte Carlo simulations are in full
agreement with the analytical trend (solid line) obtained from numerical
integration of the inverse Fourier transform (\ref{eqn:SDDint}). Right: analytical
results for the survival probability in rescaled units in the semi-infinite domain
as function of $x_0/\sqrt{t}$. In the inset the short time behaviour of $S(t|x_0)$
is reported, the universal crossover at $S\approx0.925$ is distinct.}
\label{fig:S_x0_IFFT}
\end{figure}

\begin{figure}
\centering
\includegraphics[width=0.5\textwidth]{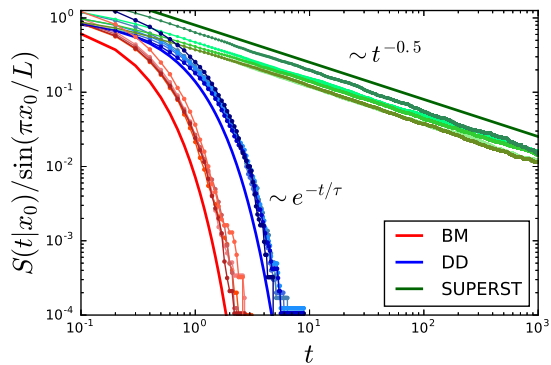}
\caption{Survival probability for the finite interval $[0,L]$, showing a comparison
between Brownian-Gaussian diffusion, DD, and superstatistical dynamics. Different
colour shades represent different initial positions $x_0$, and the solid lines
represent the analytical trends. The normalisation is chosen considering that all 3
models present the dependence on the initial position through the sinusoidal function
reported in the $y$-axis and that at long times just the first eigenstate dominates.
Indeed we observe that at long times all lines approach a quasi-master curve which
is different for each model and in agreement with the analytical results: a power law
asymptotic for the superstatistical model and exponential tails for the DD and
Brownian-Gaussian models, the latter with two different $\tau$ values of the
dominant exponential tail.}
\label{fig:S_DD_L}
\end{figure}

Figures \ref{fig:S_x0_IFFT} and \ref{fig:S_DD_L} show a comparison of the results
obtained for the DD model with the classical ones for Brownian-Gaussian motion in
the semi-infinite and finite domains, respectively. In figure \ref{fig:S_x0_IFFT}
(left) and figure \ref{fig:S_DD_L} we include results from simulations, demonstrating
excellent agreement with our analytical results. As expected, we observe significant
dissimilarities between the two models mostly in the short time limit. At intermediate
time scales the DD model shows a crossover from short time superstatistical behaviour
to the limiting Brownian-Gaussian behaviour with effective diffusivity $\langle D
\rangle_{\mathrm{st}}$.

For the semi-infinite domain figure \ref{fig:S_x0_IFFT} demonstrates that in the
short time regime the DD process exhibits a faster decay of the survival probability
and thus a more efficient first passage dynamics. This effects is particularly
visible in the right panel, in which short times correspond to large values on the
abscissa $x_0/\sqrt{t}$. To clarify this effect we express result (\ref{eqn:S_DD_ST})
and the one for Brownian motion in terms of elementary functions,
\begin{eqnarray}
S_\mathrm{BM}(t|x_0)&\sim1-\frac{\sqrt{2}e^{-(x_0^2/2t)}}{\sqrt{\pi}x_0}t^{1/2}, 
& x_0/\sqrt{t}\to\infty, \label{eqn:series_BM}\\
S_\mathrm{ST}(t|x_0)&\sim1-\frac{\sqrt{2}e^{-(x_0/\sqrt{t})}}{\sqrt{\pi x_0}}t^{1/4}
+\frac{5e^{-(x_0/\sqrt{t})}}{4\sqrt{2\pi x_0^3}}t^{3/4}, \quad & x_0/\sqrt{t}\to\infty.
\label{eqn:series_DD}
\end{eqnarray}
Comparing the asymptotes (\ref{eqn:series_BM}) with (\ref{eqn:series_DD}) along with
the inset in figure \ref{fig:S_x0_IFFT} (right), we observe that for a fixed initial
position $x_0$ the DD survival probability initially indeed drops faster than the one
for Brownian-Gaussian motion. This behaviour is more visible for larger $x_0$ and
becomes less and less relevant when $x_0$ approaches the absorbing boundary. From a
physical point of view, this can be understood due to the fact that the closer to
the boundary we place the particle initially the more likely it is that the particle
is absorbed immediately, independently from the underlying diffusive model.

Figure \ref{fig:S_x0_IFFT} (right) demonstrates two universalities. First, we observe
that at intermediate times the survival probabilities for any initial position show
a universal convergence to a common crossover point at around $S(t|x_0)\approx0.925$,
including the Brownian-Gaussian survival probability. At times shorter than this
crossover point Brownian-Gaussian motion is outperformed by the DD model, which
assumes smaller values of $S(t|x_0)$. At times longer than the crossover time the
decay of Brownian-Gaussian motion is the fastest. Second, the initial advantage of
the DD first passage dynamics over Brownian-Gaussian motion which reverts after the
universal crossover point, appears to balance out: at long times the survival
probability in all cases converges to the exact result of Brownian-Gaussian motion
with effective diffusivity $\langle D\rangle_{\mathrm{st}}$. This can be seen directly
from result (\ref{S_DD_lt}), the associated first passage density of which is
exactly the well-known L{\'e}vy-Smirnov form $\wp(t)=(x_0/\sqrt{4\pi\langle D\rangle_
{\mathrm{st}}t^3})\exp(-x_0^2/[4\langle D\rangle_{\mathrm{st}}t]$).

Qualitatively a similar behaviour is observed for finite domains at short times. As
shown in figure \ref{fig:S_DD_L}, in contrast, the long time behaviour is dominated
by the exponential shoulder (\ref{eqn:SDDL_LT}) corresponding to the lowest non-zero
eigenvalue in the DD model. The corresponding characteristic time scale $\tau$ in
figure
\ref{fig:S_DD_L} is longer than for Brownian-Gaussian motion. This is due to the fact
that in the finite interval the particles will reach the boundary before experiencing
the entire diffusivity space, and so the effective Brownian limit is not recovered.
The larger the interval $L$ is the smaller the difference between the characteristic
times of DD and Brownian-Gaussian models will be. In the limit of $L\to\infty$ the
same long time behaviour is observed. Figure \ref{fig:S_DD_L} also demonstrates an
interesting behaviour of the superstatistical model. When the diffusivity distribution
(\ref{eqn:PD}) governs the particle motion at all times $t$, even in the finite domain
a power law scaling of the survival probability emerges, and thus a diverging mean
first passage time is produced. This behaviour is caused by appreciable fraction of
immobile particles manifested in the divergence or nonzero value of $p_D(D=0)$ in
$d=1$ and $d=2$, respectively. This behaviour is rectified in the DD model.

\section{Conclusions}

We studied the first passage behaviour of the popular DD model used as a mean field
proxy for diffusion of test particles in heterogeneous environments, in which the
particle experiences varying diffusivities. Our analysis demonstrated that at
short times the DD dynamics leads to a faster decay of the survival probability and
thus to more efficient first passage. In a semi-infinite domain, fully independent
of the initial particle position a universal crossover occurs, beyond which the
DD dynamics becomes less efficient than pure Brownian-Gaussian motion, and the
ultimate decay is determined by the conventional L{\'e}vy-Smirnov behaviour for
initial particle position $x_0$ and effective diffusivity $\langle D\rangle_{\mathrm{
st}}$. The initial advantage of the DD dynamics may be particularly relevant in
cases of molecular regulation processes at very low concentrations (few-encounter
limit) \cite{aljaz}. At long times in finite domains the DD first passage
behaviour is dominated by an exponential shoulder with a characteristic time
(approximately the mean first passage time) that is longer than that for
Brownian-Gaussian motion.

These results are in agreement with the expectation that rare events, represented
by the exponential tails of the particles displacement distribution at short
times, may dominate triggered actions. Thus, even if in general heterogeneity
in the environment does not improve the mean first passage result (in fact
some of the particles are slowed down) it allows some other particles to
have a diffusion coefficient greater than the average, and this is enough
to increase the efficiency of the reaction activation.  Moreover, we proved
that the amount of fast particles is independent on the initial position,
representing the distance between particle and target. This suggests that
the obtained results may be qualitatively generalised to any distribution
of the initial particle position.

The study developed here is not limited to the one-dimensional case. First of
all, we know that in the semi-infinite domain the results of the survival
probability of Brownian-Gaussian motion in $d=2$ and $d=3$ are the same as
the one in $d=1$. Then, the same analysis of the first passage problem can be
performed by solely changing to the corresponding $d$-dimensional subordinator. For finite
domains the analysis is also similar since, for all $d$ we have an exponential
behaviour in time of the propagator which allows us to relate the DD survival
probability to the Laplace transform of the corresponding subordinator,
as we did for the one-dimensional case.

We finally note that similar non-Gaussian effects have been reported for systems,
in which the (subdiffusive) motion is dominated by viscoelastic effects. With a
fixed diffusivity this would be a Gaussian process, and the non-Gaussianity was
shown to stem from varying diffusivity values \cite{prx,Lampo:BYNG6,natcomm}.
It will be interesting to study the associated first passage behaviour in this
case, as well.

\ack

VS thanks the organisers of the 671st WE-Heraeus-Seminar on Search and Problem Solving
by Random Walks: Drunkards vs Quantum Computers at Bad Honnef, Germany, in May 2018
where the main results of this work were presented. RM and AVC acknowledge funding
from Deutsche Forschungsgemeinschaft, grants ME 1535/6-1 and ME 1535/7-1. RM
acknowledges the Foundation for Polish Science (Fundacja na rzecz Nauki Polski) for
funding within an Alexander von Humboldt Polish Honorary Research Scholarship.

\section*{References}

\bibliographystyle{iopart-num}

\end{document}